\documentclass[nocompress]{spie}

\usepackage{colordvi}
\usepackage{graphicx}
\usepackage{hyperref}
\usepackage{url}

\title{The Thermal Environment of the Fiber Glass Dome for the New Solar
    Telescope at Big Bear Solar Observatory}
\author{A.~P.\ Verdoni\supit{a}, C.\ Denker\supit{a,b}, J.~R.\ Varsik\supit{c},
    S.\ Shumko\supit{c}, J.\ Nenow\supit{c}, and R.\ Coulter\supit{c}
\skiplinehalf
\supit{a}New Jersey Institute of Technology, 323 Martin Luther King Blvd,
    Newark, NJ~07102, U.S.A.\\
\supit{b}Astrophysikalisches Institut Potsdam, An der Sternwarte 16,
    D-14482 Potsdam, Germany\\
\supit{c}Big Bear Solar Observatory, 40386 North Shore Lane, Big Bear City,
    CA~92314, U.S.A.}

\begin{document}
\maketitle

%
%

\begin{abstract}
The New Solar Telescope (NST) is a 1.6-meter off-axis Gregory-type telescope
with an equatorial mount and an open optical support structure. To mitigate the
temperature fluctuations along the exposed optical path, the effects of
local/dome-related seeing have to be minimized. To accomplish this, NST will be
housed in a 5/8-sphere fiberglass dome that is outfitted with 14 active vents
evenly spaced around its perimeter. The 14 vents house louvers that open and
close independently of one another to regulate and direct the passage of air
through the dome. In January 2006, 16 thermal probes were installed throughout
the dome and the temperature distribution was measured. The measurements
confirmed the existence of a strong thermal gradient on the order of $5^{\circ}$
Celsius inside the dome. In December 2006, a second set of temperature
measurements were made using different louver configurations. In this study, we
present the results of these measurements along with their integration into the
thermal control system (ThCS) and the overall telescope control system (TCS).
\end{abstract}


%
%

\section{INTRODUCTION}

Solar telescopes with an aperture larger than 1~meter face a variety of
challenges, if they want to achieve diffraction-limited resolution. Site
selection is of primary concern. In the context of the proposed 4-meter aperture
Advanced Technology Solar Telescope (ATST), significant efforts were undertaken
to identify the best site(s) for solar observations. \cite{Hill2004,
SocasNavarro2005, Verdoni2007} Big Bear Solar Observatory (BBSO) was identified
as one of three sites suitable for high-resolution solar observations. However,
the seeing characteristics at BBSO -- a mountain-lake site -- are quite
different \cite{Verdoni2007} from the two other mountain-island sites
Haleakal\={a}, Maui, Hawai'i (which was selected as the future ATST site) and
Observatorio Roque de los Muchachos on La Palma, Spain. The lake effectively
suppresses the ground-layer seeing and very good seeing conditions are
encountered from sunrise to sunset. This makes BBSO ideally suited for solar
activity monitoring and space weather studies \cite{Gallagher2002} combining
synoptic with high-resolution capabilities. These site characteristics and
scientific objectives are exactly what has motivated the design, development and
now construction of NST. \cite{Goode2003a, Denker2006b}

Since instrument seeing is a severe issue for solar telescopes, most
high-resolution solar telescopes were placed inside vacuum tanks. This approach,
however, is no longer feasible for apertures larger than 1~meter, since the
entrance window (or lens) would become too thick in order to withstand the
vacuum. Therefore, the next generation of solar telescopes has to rely on
``open-designs'', i.e., the optical support structure and optics will be exposed
to the elements. This in turn requires a good understanding and control of the
thermal environment in which the telescope is placed. In a first set of
papers,\cite{Denker2006a, Verdoni2006} we have described on how to integrate
seeing measurements into the NST operations and introduced plans on how to
implement the NST ThCS. In this study, we will discuss more details on the ThCS
implementation, the design of the fiberglass dome (a smaller version of the
SOuthern Astrophysical Research (SOAR) telescope dome \cite{Teran2000}, a
detailed weather record for BBSO, temperature measurements inside the dome under
varying observing conditions, and some implications for the thermal control of
the primary mirror, which is a 1/5-scale model for the 8.4-meter off-axis
segments of the Giant Magellan Telescope's primary mirrors.\cite{Martin2004}

%
%

\section{5/8-SPHERE FIBERGLASS DOME AND LOUVER EXPERIMENTS}

The new NST dome at BBSO is a 10-meter diameter 5/8 sphere with an over the top
nesting shutter housing a 2-meter circular aperture. The dome was  manufactured
by MFG Ratech and is modeled after the dome for the SOAR  telescope in Cerro
Pachon, Chile.\cite{Teran2000} The SOAR dome is approximately twice the size of
the NST with a diameter of 20~m. It has a similar over the top shutter with
windscreen attachment. However, it does not have the 14 active damped louver
assemblies, which are evenly spaced around NST's equator. The exterior of the
NST dome is comprised of Fiberglass Reinforced Plastic (FRP) panels that are
assembled in three ring sections, which are vertically split into two
hemispheres by the dome slit. The sections are supported by two 10-meter
diameter steel arch girders that serve as guides for the dome shutter. Both the
panels and arches sit on top of a 9.2-meter diameter steel ring beam, which
rotates on a 16 fixed-point bogie system allowing the dome, shutter and
windscreen to track the telescope for maximum protection against the prevailing
winds.

\begin{figure}[t]
\includegraphics[bb=0 0 800 533,width=84mm]{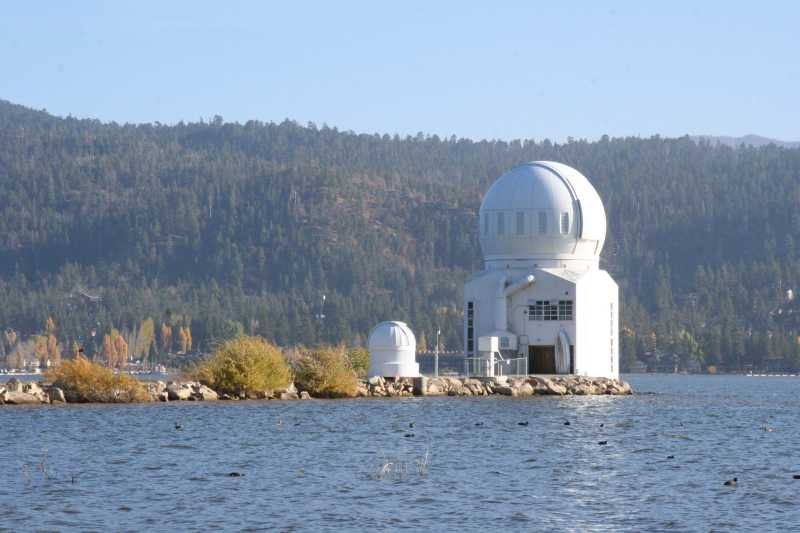}
\hfill
\includegraphics[bb=0 0 800 533,width=84mm]{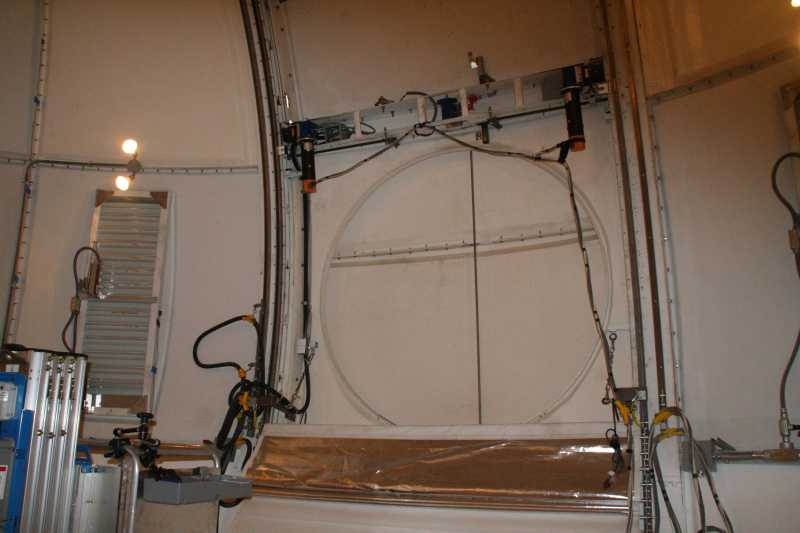}
\caption{{\it Left.} Image of the new NST dome at BBSO. The smaller dome in the
    foreground houses the Earthshine and H$\alpha$ full-disk telescopes. In this
    image the shutter is closed and the iris is covering the domes 2-meter
    circular aperture. {\it Right.} Inside of view of the closed iris and the
    iris drive motors. The folded windscreen is visible at the bottom of the
    iris.}
\label{FIG01}
\end{figure}

The left panel of Figure~\ref{FIG01} shows the newly installed NST dome with the
aperture pointing to the east. In this image the dome shutter is closed and the
iris covers the 2-meter circular aperture. Housed in the smaller dome in the
foreground are the BBSO Earthshine and H$\alpha$ full-disk telescopes. The right
panel of Figure~\ref{FIG01} depicts an inside view of  the closed iris. At the
bottom of the aperture panel is the collapsed, foldable wind screen. As the
aperture panel raises and lowers, the wind screen unfolds covering the exposed
portion of the dome slit. To test the structural integrity of the dome a stress
analysis was performed. The analysis was based on the maximum operating snow and
ice loading conditions. The conditions call for the dome to retain its structure
with a snow depth of 1~m, an average dead load of 97~kg~ft$^{2}$ with a peak
dead load of 195~kg~ft$^{2}$, and an ice thickness of 0.05~m. A worst case
stress analysis was performed with a peak dead load of 195~kg~ft$^{2}$ and a
200~kph wind acting simultaneously. The result was that the 10~m diameter was in
substantial conformance with the manufacturers requirements.

NST's 14 vent gates allow wind flushing of the dome interior. The gates are made
of a heavy gauge extruded aluminum, which is rated to withstand a wind load of
approximately 190~kg~m$^{-2}$. Each of the vents is 0.6~m$\times$1.8~m with a
depth of approximately 0.1~m. A damper is attached to each vent, which allows
control over the amount and direction of air flow through the dome. The dampers
and will be activated based on the direction of the wind, measured by a weather
station outside the dome, and the temperature gradients inside the dome,
measured by 16 temperature probe units arranged symmetrically throughout the
dome interior.\cite{Denker2006b}

\begin{figure}[t]
\includegraphics[bb=0 0 800 533,height=60mm]{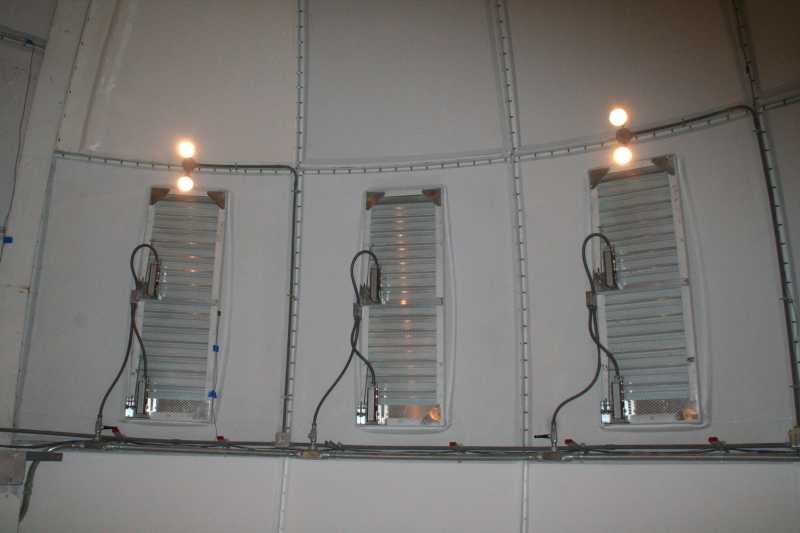}
\hfill
\includegraphics[bb=0 0 800 600,height=60mm]{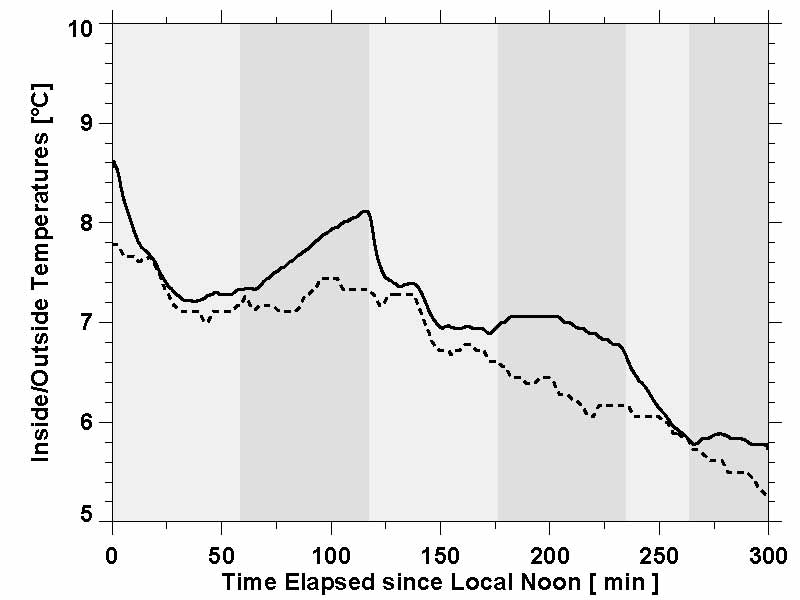}
\caption{{\it Left.} The inside of the dome showing three of the active vent
    gate assemblies. Visible on each of the vents are the two motors that adjust
    the damper settings. {\it Right.} Temperature measured by one thermal probe
    (solid line) inside of the dome and the outside temperature (dashed line) as
    recorded by the weather station. The alternating light and dark stripes
    correspond to the open and closed configurations of the 14 louvers,
    respectively.}
\label{FIG02}
\end{figure}

The left panel of Figure~\ref{FIG02} is an image of the inside of the dome
showing three vent gate assemblies. The two motors responsible for opening and
closing the vent louver system. To eliminate the presence of a thermal gradient
inside the dome, the dome volume must be flushed 20 or more times per hour,
which requires a wind speed of approximately 2 to 3~m~s$^{-1}$. BBSO benefits
from a predominately westerly winds with a mean speed of
6~m~s$^{-1}$.\cite{Verdoni2007} With a total dome volume of 330~m$^{3}$, this is
sufficient to achieve a flushing rate of 30 dome volumes per hour.

In January 2006, the first set of 16 temperature sensors were installed on the
interior surface of the NST dome. The dome structure, skin, and louvers were
installed at this time. However, the dome drive and shutter motors were not yet
operational. The temporal evolution of the dome temperature was measured
throughout the course of the day showing the existence of thermal gradients on
the order of $10^{\circ}$~C or more. In December 2006, with control of the dome
drive and shutter motors being available, we were able to monitor the temporal
evolution of the dome temperature while simultaneously opening and closing the
louvers.

The dome shutter was closed as well as was the iris. The aperture was pointing
to the west and remained in this position for the duration of the experiment. To
control the opening and closing of the louvers the positions were set manually.
The probe temperatures were obtained once every minute by polling the Temp-Trax
Model E16 tracking thermometer using a Perl script running on a BBSO server. The
configurations for the louvers were decided upon a priori and then executed.
Along with the thermal probe data the corresponding weather station data was
downloaded from the weather station.

The right panel of Figure~\ref{FIG02} shows the results  of a louver experiment
conducted on December 12, 2006. The $y$-axis is temperature in  degrees Celsius
and the $x$-axis displays the minutes elapsed since local noon. The dark and
light stripes of the background correspond to all 14 louvers being open or
closed, respectively. The solid line is the temperature measured by a single
thermal probe and the dashed line is the outside temperature measured by the
weather station. The louvers are opened and closed every hour for the first four
hours, then opened for a half hour and closed ending the run. The objective will
be to determine with a good degree of accuracy the response time of the internal
dome temperature with respect to louver configuration. Looking at the  plot one
can immediately see that the temperature inside of the dome follows the general
trend of the exterior temperature. The opening and closing of the louvers is
also apparent. When the louvers are open the temperature shows a characteristic
decline until the interior and exterior are in close agreement. When the louvers
are closed the temperature inside the dome climbs to approximately 1$^{\circ}$~C
above the outside temperature. For more accurate results these experiments
should be carried out  with the NST present to ensure that all thermal sources
are accounted for.

\begin{figure}[t]
\includegraphics[bb=0 0 800 600,width=84mm]{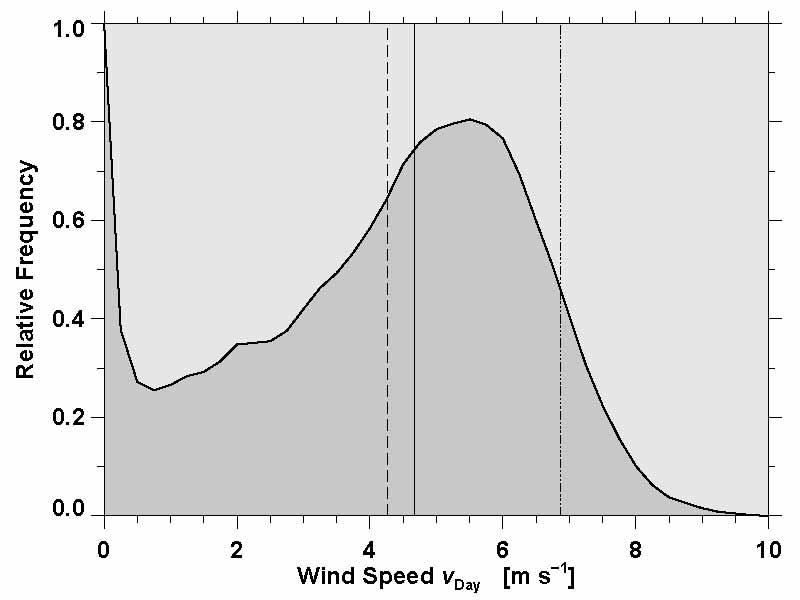}
\hfill
\includegraphics[bb=0 0 800 600,width=84mm]{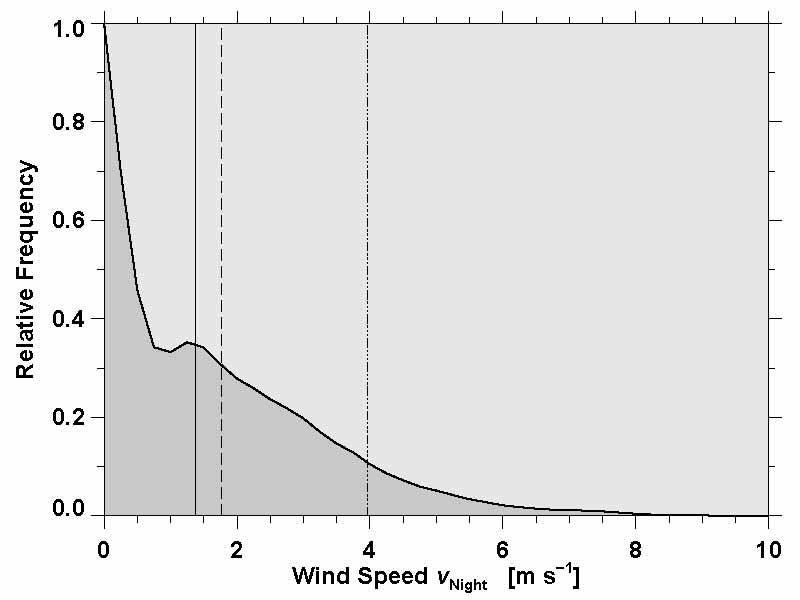}
\caption{Frequency distributions of the wind speeds $v_{\rm Day}$ ({\it left})
    and $v_{\rm Night}$ ({\it right}). The median, mean, and 10th percentile
    wind speeds are indicated by solid, dashed, and dashed-dotted vertical
    lines, respectively.}
\label{FIG03}
\end{figure}

\begin{figure}
\includegraphics[bb=0 0 800 600,width=84mm]{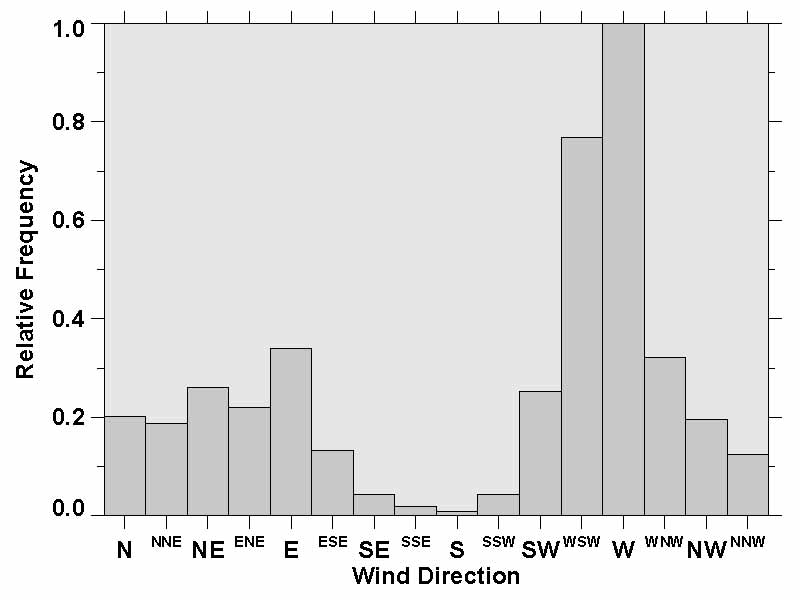}
\hfill
\includegraphics[bb=0 0 800 600,width=84mm]{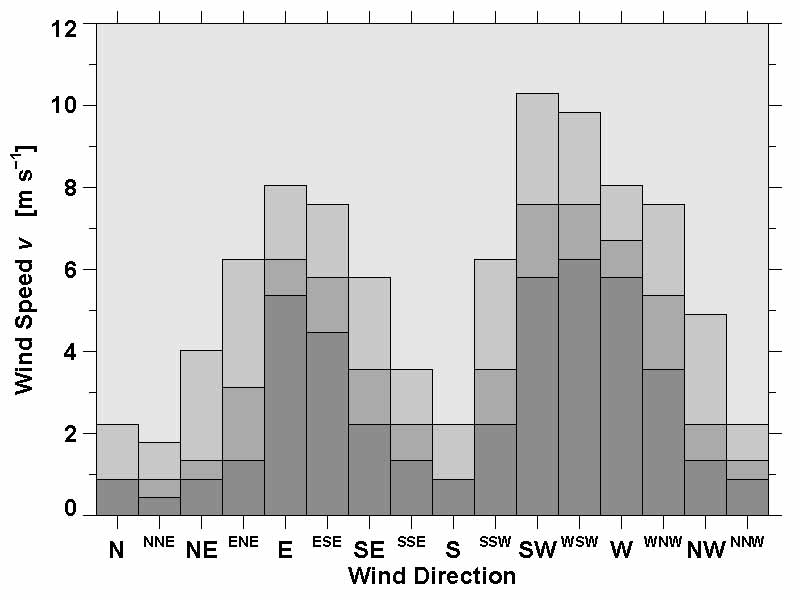}
\caption{{\it Left.} Frequency distribution of the wind directions.
    {\it Right.} The median wind speed as a function of the wind direction is
    shown as dark gray. The lighter grays correspond to the 10th and 30th
    percentile of the respective frequency distributions.}
\label{FIG04}
\end{figure}

%
%

\section{METEOROLOGICAL DATA}

An accurate characterization of the meteorological conditions in the immediate
surroundings of an observatory is important for daily and seasonal operations of
the telescope. In February 2005, a Vantage Pro2 Plus weather station
manufactured by Davis Instruments (\url{http://www.davisnet.com/}) was installed
at BBSO to monitor these conditions. The weather station is outfitted with an
integrated suite of meteorological instruments including solar radiation and UV
sensors. Temperature and humidity sensors are housed inside a radiation shield
for improved accuracy. The weather station data presented in this section covers
741 days from February~28, 2005 to March~10, 2007. The data was sampled at a
1-minute cadence with the exception of the first 10 days, when a cadence was
5~min.

The frequency distributions of daytime ({\it left}) and nighttime ({\it right})
wind speeds are shown in Figure~\ref{FIG03}. The median wind speed during the
daytime is 4.67~m~s$^{-1}$ with a mean of 4.27~m~s$^{-1}$. The 10th percentile
(6.87~m~s$^{-1}$) of the wind speed distribution was computed to provide an
estimate for high wind conditions. During the night, the winds decrease in
strength. The median, mean, and 10th percentile wind speeds are 1.38~m~s$^{-1}$,
1.76~m~s$^{-1}$, and 3.97~m~s$^{-1}$, respectively. The median, mean and 10th
percentile values for each of the distributions are represented by a solid,
dashed, and dashed-dotted vertical line in Figure~\ref{FIG03}. The two wind
speed distributions clearly show different characteristics. Much higher
velocities are encountered during the daytime with a well-defined maximum
between 5.0 and 6.0~m~s$^{-1}$. This maximum is basically absent in the
nighttime distribution, where we find a basically monotonic decrease of the
frequency of occurrence with wind speed. The physical mechanism behind these
discrepancies becomes more apparent in the frequency distribution of the
directions (Figures~\ref{FIG04}).

\begin{figure}[t]
\includegraphics[bb=0 0 800 600,width=84mm]{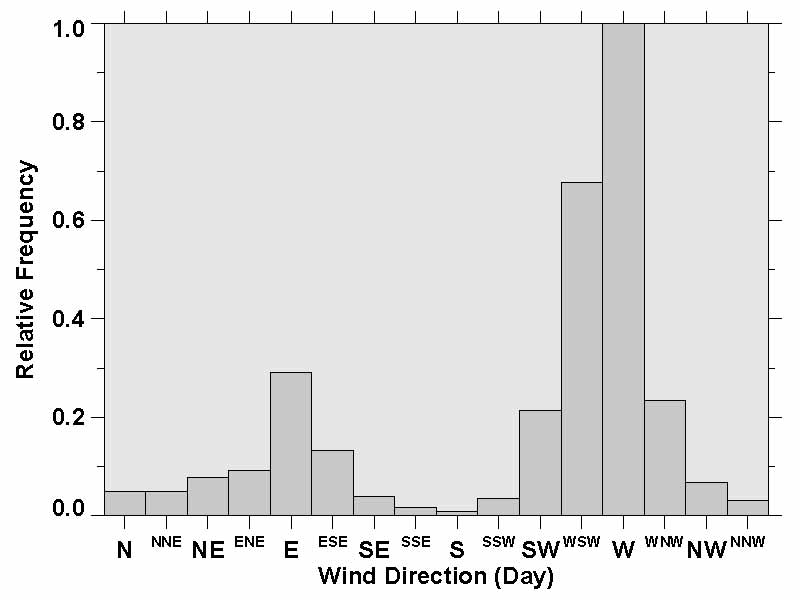}
\hfill
\includegraphics[bb=0 0 800 600,width=84mm]{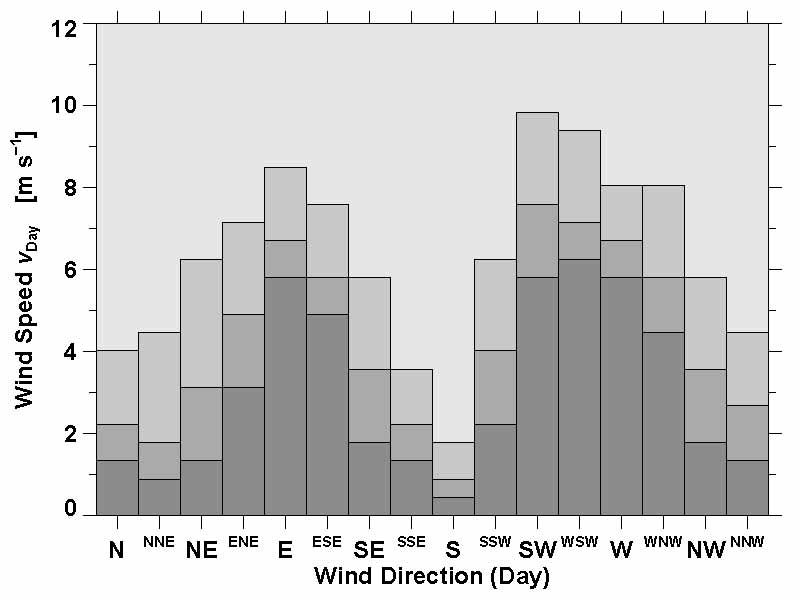}
\caption{Frequency distribution of the daytime wind directions (same format as
    Figure~\ref{FIG04}).}
\label{FIG05}
\end{figure}

\begin{figure}
\includegraphics[bb=0 0 800 600,width=84mm]{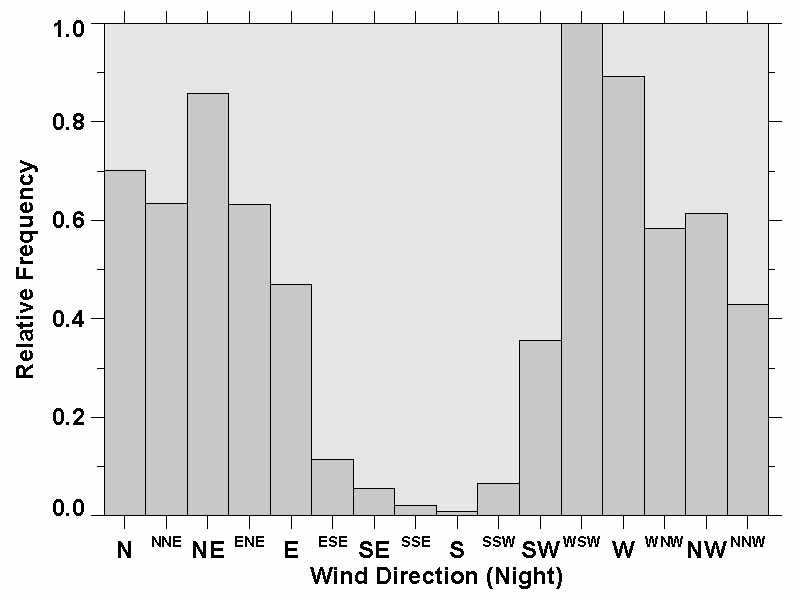}
\hfill
\includegraphics[bb=0 0 800 600,width=84mm]{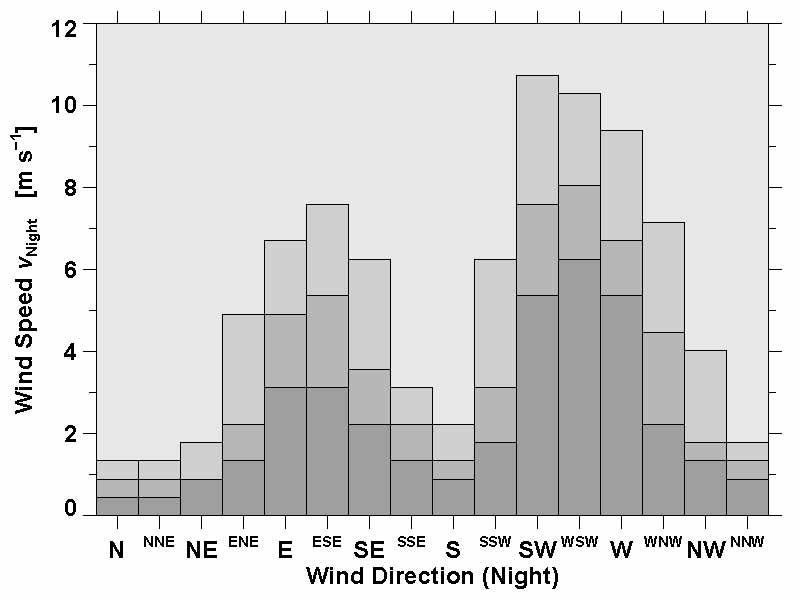}
\caption{Frequency distribution of the nighttime wind directions (same format as
    Figure~\ref{FIG04}).}
\label{FIG06}
\end{figure}

The east-west orientation of Big Bear Lake and its mountain location gives rise
to a unique distribution of wind directions. The relative frequency distribution
of the wind direction is shown in the left panel of Figure~\ref{FIG06} for the
entire data set. The wind directions in the following plots are given on a the
compass rose graduated into 16 sector. Immediately apparent is the predominance
of westerly wind directions. This westerly wind is a result of gradient in
pressure between the Los Angeles basing and the interior regions of the Southern
Californian desserts.\cite{Verdoni2007} The winds are channeled through the
canyons an valleys of the San Bernardino mountains. Since the wind does not
encounter any obstruction passing over the cold waters of Big Bear Lake, the
observatory island is embedded in almost laminar air flows. The water also
provides a ``heat sink'' effectively suppressing ground-layer seeing. This is
the explanation for the very good seeing conditions from sunrise to sunset at
BBSO. A more quantitative picture of the wind speeds as a function of wind
direction is presented in the right panel of Figure~\ref{FIG04}. Here, the gray
scale corresponds (from dark to light) to the median, 10th and 30th percentile
of the frequency distributions, respectively. For a westerly wind a median wind
speed of approximately 6.0~m~s$^{-1}$ is measured with winds of 8~m~s$^{-1}$ or
greater occurring 30 percent of the time. For easterly wind directions, the
speeds are comparable to westerly winds. These winds are caused by a reversal of
the pressure gradient and are commonly referred to as Santa Ana
winds.\cite{Hu2003} A detailed discussion of the daytime seeing characteristics
at BBSO was presented in an earlier study\cite{Verdoni2007} based on data from
the ATST site survey. However, due to the nature of the daytime seeing monitor
no nighttime weather data was available. Since nighttime observations might be
scheduled for NST and BBSO has an existing program for Earthshine observations,
\cite{MontanesRodriguez2005} we present in this study also the respective
nighttime weather data. These data also further illustrate the two different
observing regimes for day and night (Figures~\ref{FIG05} and \ref{FIG06}).

\begin{figure}
\includegraphics[bb=0 0 800 600,width=84mm]{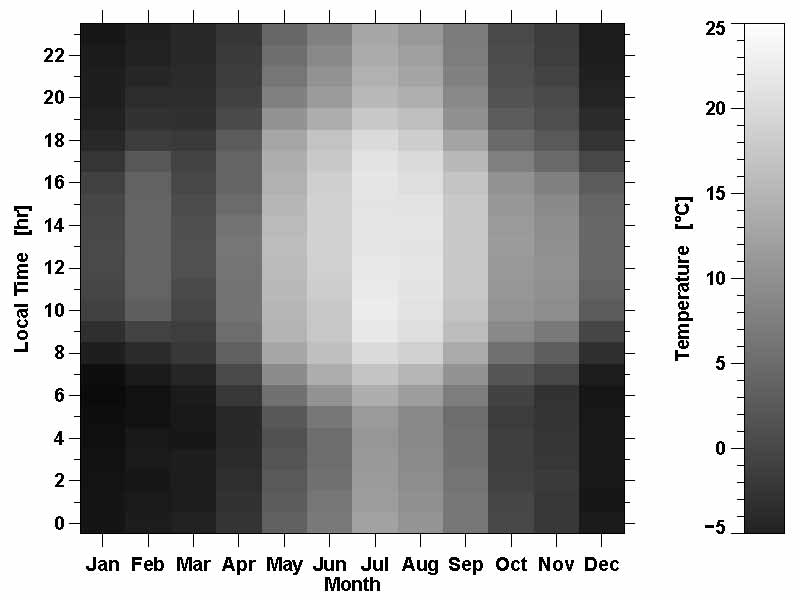}
\hfill
\includegraphics[bb=0 0 800 600,width=84mm]{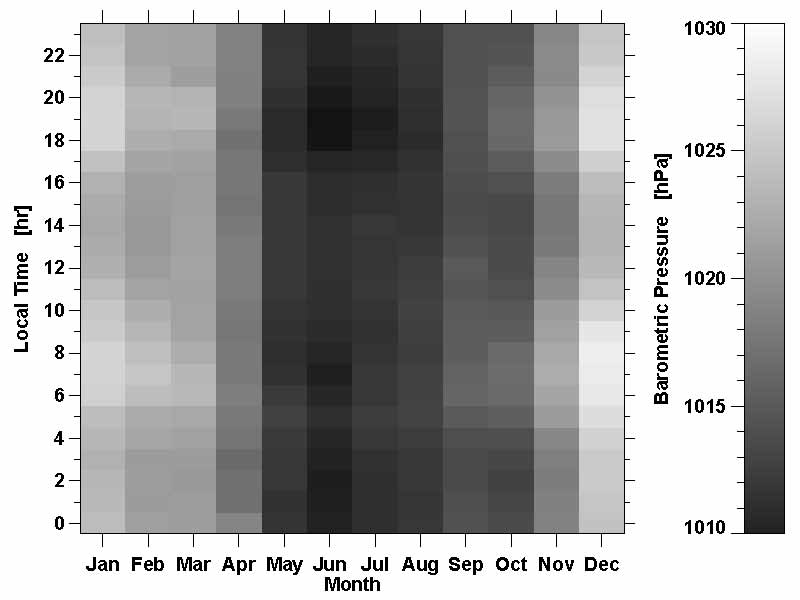}
\caption{{\it Left.} Seasonal and diurnal variation of median temperature.
   {\it Right.} Seasonal and diurnal variation of median barometric pressure.}
\label{FIG07}
\end{figure}

\begin{figure}[t]
\includegraphics[bb=0 0 800 600,width=84mm]{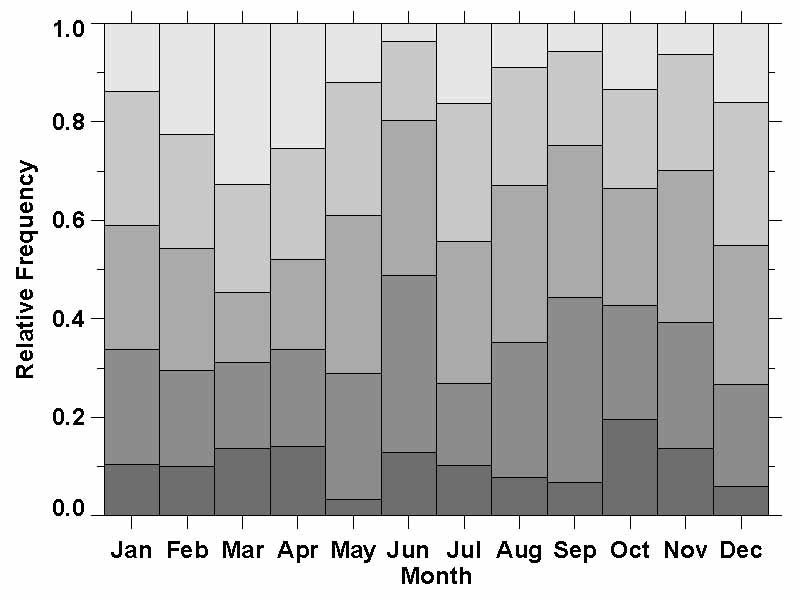}
\hfill
\includegraphics[bb=0 0 800 600,width=84mm]{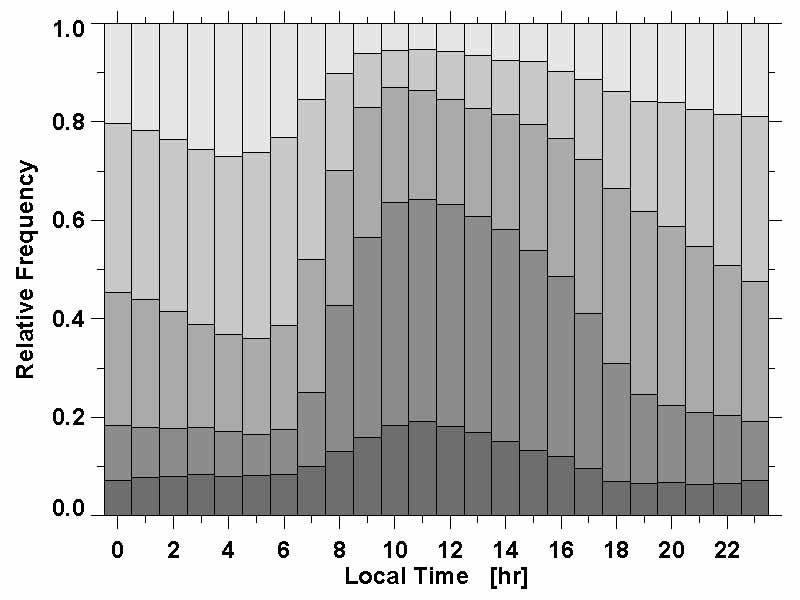}
\caption{{\it Left.} Seasonal frequency distributions of the humidity.
    {\it Right.} Diurnal frequency distributions of the humidity. The
    gray scale (from dark to bright) corresponds to humidity levels of 20\%,
    40\%, 60\%, and 80\%.}
\label{FIG08}
\end{figure}

Separating the daytime and nighttime frequency distributions shows an even more
pronounced east-west orientation of the winds during the day. It also shows that
Santa Ana conditions do not very frequently occur (about 10\% of the time).
However, even under these conditions the seeing can be very
good\cite{Verdoni2007} but the observing conditions suffer from a low sky
transparency due to dust carried in from the deserts. The wind speeds are very
similar (about 6.0~m~s$^{-1}$) for winds from the East and West (see right panel
in Figure~\ref{FIG05}. However, in the the North-South direction the wind speeds
are reduced to about 1.5~m~s$^{-1}$. Figure~\ref{FIG06} for reveals the
nighttime wind regime, which is dominated by mountain down-slope winds. The air
that has been heated during the day, now slowly flows down the mountain slope
towards the cool surface of the lake. With the exception of the southern
mountain slopes, there is no preferential direction for the down-slope winds but
the wind speed distribution again follows the east-west orientation of Big Bear
Valley. Consulting the wind speed distributions in Figure~\ref{FIG03}, low wind
speed times $v < 1.0$~m~s$^{-1}$ occur during a significant fraction of time.
This happens typically around dusk and dawn, when the change between the daytime
and nighttime wind regime takes place.

\begin{figure}[t]
\includegraphics[bb=0 0 768 576,height=55mm]{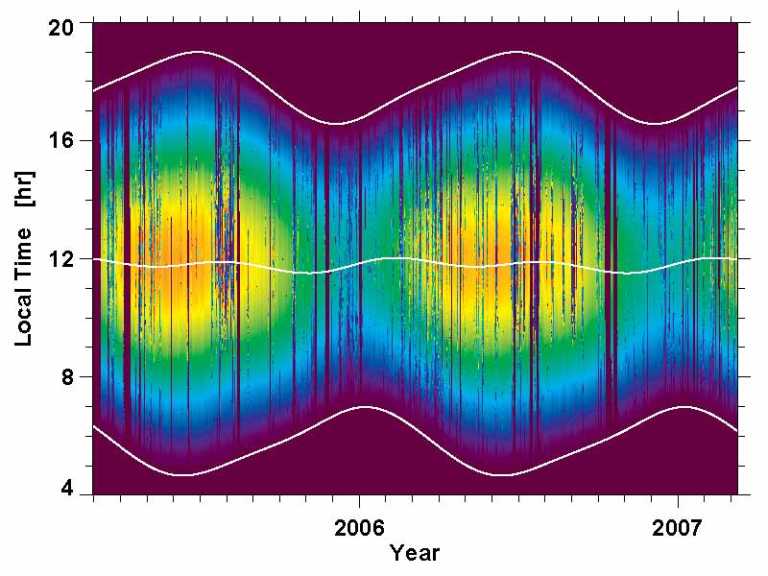}
\hfill
\includegraphics[bb=0 0 612 387,height=55mm]{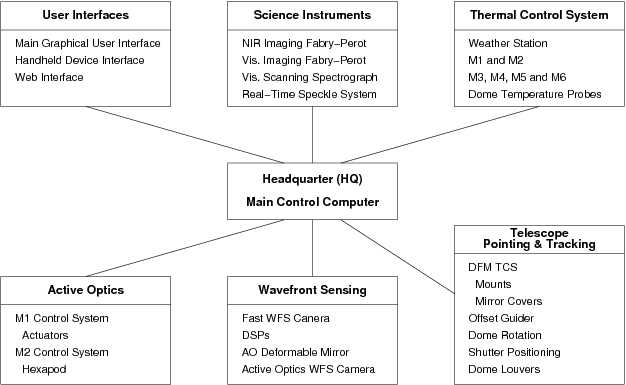}
\caption{{\it Left.} Diurnal and seasonal variation of the solar radiation
    spanning the entire data set. The white contour lines refer to sunrise,
    local noon, and sunset. {\it Right.} Schematic overview of the main TCS
    components. The system components of the ThCS and TPTS, which are
    responsible for controlling the NST thermal environment, are shown on the
    right.}
\label{FIG09}
\end{figure}

In Figure~\ref{FIG07}, the seasonal and diurnal variations of the median
temperature ({\it left}) and median barometric pressure ({\it right}) are shown
as gray scales. The hottest month is July, when median temperatures reach about
$25^{\circ}$~C. The temperature spread between day and night is also much large
during the summer months (about $10^{\circ}$~C compared to $7^{\circ}$~C in the
winter). The warm temperatures in February might be an anomaly, since our data
sample only covers slightly more than two years. The barometric pressure shown
in the right panel of Figure~\ref{FIG07} is basically inversely proportional to
the temperature. The lowest values of the barometric pressure are measured
throughout the summer months. The diurnal variations are more or less
negligible. However, a small trend to higher barometric  pressure can be
observed around dusk and dawn, when transitioning from the daytime to the
nighttime wind regime and vice versa. The seasonal trend of the barometric
pressure is given by the pressure gradient between the coastal regions of
Southern California and the inland deserts.

The seasonal ({\it left}) and diurnal ({\it right}) distributions of the
humidity are shown in Figure~\ref{FIG08}. The gray scale corresponds (from dark
to bright) to humidity levels of 20\%, 40\%, 60\%, and 80\%, respectively.  The
winter months and early spring is the most humid time of the year. Surprisingly,
only a weak indication is found (in July) is visible for the monsoon season in
July and August. Again, the explanation for this behavior is in the  diurnal
changes of the humidity. Shortly after sunrise, the air dries out and becomes
least humid around local noon. With decreasing solar input in the afternoon, the
humidity increases monotonically. This trend continues after sunset. The most
humid time is reached just before sunrise. The high humidity during the night
and dawn might cause problems for nighttime and Earthshine observations.

The weather station also included a solar radiometer. The left panel of
Figure~\ref{FIG09} displays the diurnal and seasonal variations of the solar
radiation from which the clear time fraction (CTF) was computed. The solar
radiation is displayed on a rainbow color scale, where violet corresponds to low
light levels and orange/red refers the maximum of the solar radiation at noon in
the summer time. Solar ephemeris computations including optical air mass were
obtained from the Jet Propulsion Laboratory (JPL) Horizons web page
(\url{http://ssd.jpl.nasa.gov/horizons.cgi}) to aid in the
analysis.\cite{Giorgini1996} The ephemeris data were converted from Universal
Time (UT) to Pacific Standard Time (PST, without daylight savings time
correction) to match the weather station data. The three white lines (from
bottom to top) refer to sunrise, local noon, and sunset, respectively.
Especially in the winter and spring, entire days with low light levels are
recorded. These times are during winter storms and when the sky is overcast. The
monsoon season is also visible in the left panel of Figure~\ref{FIG09}during the
months of July and August, when the daily radiation traces become spotty about
two hours before local noon. During the monsoon the moist air and solar heating
can give rise to severe thunderstorms. A threshold of 90\% of the instantaneous
solar radiation is used to compute the CTF. Since this fractional criterion
becomes impractical for elevations lower than $5^{\circ}$, we only computed the
CTF for higher elevation angles and extrapolated this value to the entire data
set. The CTF determined from the solar radiation sensor of the weather station
is 72.4\%.  This value agrees well with CTF value from the ATST site survey
(71.2\%) \cite{Hill2004} and previous measurements obtained as part of the GONG
site survey (70.7\%).\cite{Hill1994}

%
%

\section{THERMAL CONTROL SYSTEM}

The Thermal Control System (ThCS)\cite{Verdoni2006} is part of a distributed
computer system, which controls the telescope, dome, adaptive optics (AO), and
the post-focus instrumentation. The overall system is known as the Telescope
Control System (TCS).\cite{Yang2006a} A schematic overview of the TCS is shown
in the right panel of Figure~\ref{FIG09}. Use of a distributed system allows for
greater flexibility, and ultimately for greater simplicity than a monolithic
design would provide. As a consequence, each subsystem within the TCS has only a
limited set of tasks to perform.

Overall management of the TCS is carried out by the HQ (Headquarters) program
running on a dedicated main control computer. HQ collects data from each
subsystem for centralized logging and access by the main user GUI systems.
Commands are sent from the GUIs to HQ for dispatch to the appropriate TCS system
for execution. Each subsystem has an engineering GUI written in Java. In
general, C$++$ and Java are the only languages used in the control system. The
object-oriented TCS design has so far resulted in a successful implementation of
all subsystems and significantly shortened the design and development
time\cite{Shumko2006}. Even though status information is provided by HQ to
science instruments, they are currently not considered part of TCS. Instrument
designers have to rely on well-defined interfaces implemented in the eXtended
Markup Language (XML) to integrate the post-focus instruments in the
hierarchical TCS infrastructure.

Communication within TCS is performed by  Ethernet, which offers much better
performance compared to older RS-232-based implementations and simplifies
cabling to the telescope. Internet Communication Engine (ICE) from ZeroC was
chosen as the standard communication protocol/software library. User interaction
with all subsystems is handled by a system-wide main GUI, which accesses the
subsystems through HQ. The HQ computer is also performing logging and archiving
of status information, which is collected in a data base.

All high-level commands are written in XML. Commands, information requests,
notifications, or error messages can be sent asynchronously. Fast network
connectivity and generally short messages make the overhead of ASCII formatting
negligible and does not result in any significant impact on overall performance.
Processing of XML messages is trivial, since there are many ready-to-use
libraries in both Java and C$++$. ICE also allows  seamless integration various
operating systems (e.g., Windows XP and Linux), which are installed on the
various subsystem control computers. Subsystems access and control hardware
directly or through off-the-shelf Ethernet-based controllers. For example, Galil
multi-axis controllers are used with the servo motors for dome rotation and dome
shutter operation. Another, example are the TempTrax Model E interfaces for the
temperature probes throughout the dome, which communicate via a built-in web
server.

While ThCS monitors temperatures within TCS, ThCS also closely interacts with
the Telescope Pointing and Tracking System (TPTS)\cite{Varsik2006}, which
handles the movement of the telescope and dome. TPTS also provides a wrapper for
the telescope mount software provided by the telescope manufacturer DFM
Engineering, Inc. One of the TPTS challenges is the alignment of the relatively
small dome opening with the optical axis of the telescope. An algorithm is used
to find the position required for the dome so that the aperture is in the
correct place aligned with the telescope beam, allowing for the offsets between
the pivot point of the telescope mount, the center of the telescope light path,
and the center of the sphere of the dome. The dome shutter and azimuth drives
are then moved so the dome aperture follows the position of the light path
during the day. TPTS also controls the dome louvers.

The primary ThCS objective is to provide a stable environment for NST's main
optical components, which are located inside an open telescope support
structure. Therefore, the temperature of the optical support structure and
primary and secondary mirrors has to be closely monitored. The surface of the
primary mirror can be actively cooled by regulating the air flow and temperature
to a fan driven heat exchanger. The exchanger will provide a laminar flow across
the mirror surface, thus avoiding the rise of turbulent eddies from the  sunlit
mirror surface. The only way to keep the temperature of the telescope structure
and other optical components close to ambient is to effectively ventilate the
dome and limit the amount of sunlight entering through the dome aperture by
making it as small as possible. The radius of the dome aperture is only 20~cm
larger than that of the primary mirror. To accomplish the first task, ThCS
automatically sends commands to operate the fourteen dome louvers through TPTS
to adjust to variable wind speeds and directions during the observing day.
Real-time information from a weather station and from a network of temperature
probes inside the dome are used in a decision module,\cite{Denker2006a,
Verdoni2006} which sends the appropriate adjustments through HQ to the dome
louvers.

\begin{figure}[t]
\includegraphics[bb=0 0 800 600,width=84mm]{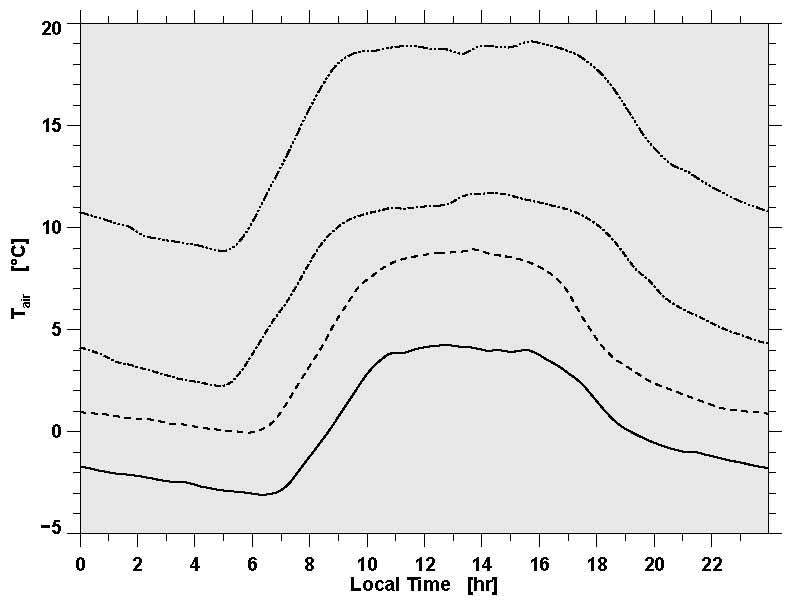}
\hfill
\includegraphics[bb=0 0 800 600,width=84mm]{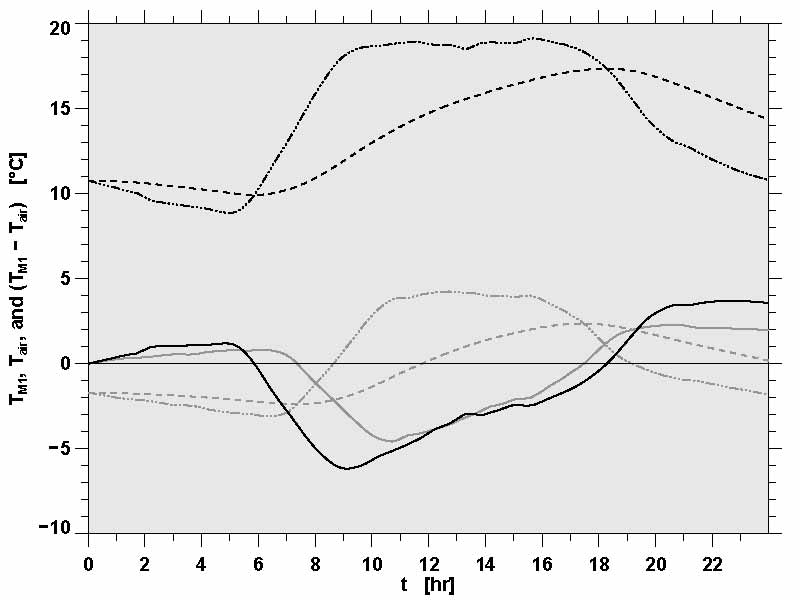}
\caption{{\it Left.} Average daily temperature profiles for winter (solid),
    spring (dashed-dotted), summer (dashed-triple-dotted), and autumn (dashed).
    {\it Right.} M1 response (dashed) to ambient air temperature variation
    (dashed-triple-dotted) during the summer (black) and winter (gray)
    considering convective heat exchange only. Initially (at midnight), M1 and
    ambient air temperature are the same. Their temperature differences are
    given by the solid curves.}
\label{FIG10}
\end{figure}

%
%

\section{THERMAL CONTROL OF THE PRIMARY MIRROR}

The thermal control of the primary mirror relies on two different mechanisms:
the air inside the dome has to track the outside temperature and a fan driven
heat exchanger directs a laminar flow of  air across the Sun-facing primary
mirror surface. Assuming that the passive ventilation through the the louvers
minimizes interior temperature gradients and reduces internal dome seeing, we
can use sample temperature profiles for winter, spring, and summer to evaluate
the thermal properties of the primary mirror under realistic observing
conditions.

Sample profiles for the four seasons are shown in the left panel of
Figure~\ref{FIG10}. These samples were created by averaging all available
temperature profiles for the respective seasons. The general shape of all
profiles is roughly the same. A fast monotonic temperature rise after sunrise,
which reaches a plateau about two hours before local noon. At this time, laminar
wind flow across the lake and the lake acting as a heat reservoir balance solar
heating and an equilibrium is reached. This plateau persists for up to 10 hours
in the summer but lasts only about six hours in the winter. This region of
little or no change is not an artifact of the averaging procedure but can also
be found in individual daily temperature profiles. Exceptions are days with
extended cloud cover. However, since BBSO has more than 300 sunny days per year,
cloudy or rainy days only leave an negligible imprint on the averaged seasonal
profiles. In the late afternoon and early evening radiative cooling begins, once
the Sun sets behind the mountains to the West of the observatory, reaching the
coolest temperatures just before sunrise. Typical temperature spreads between
day and night are 7.3~K, 9.4~K, 10.3~K, and 9.0~K for winter, spring, summer,
and autumn, respectively.

In the right panel of Figure~\ref{FIG10}, we have chosen two temperature
profiles (dashed-triple-dotted curves) for winter (gray) and summer (black) to
illustrate how the Zerodur primary responds to changes of the ambient
temperature (dashed curve). Even without direct exposure to sunlight, the
thermal inertia of primary mirror leads to a rapid departure from the ambient
temperature. Once the temperature difference exceeds about $\pm 1$~K, mirror
seeing becomes an issue and severely limits the performance of a (solar)
telescope.

In this introductory example, we assumed that M1 and the surrounding air had the
same temperature at midnight. M1 then slowly follows the cooling trend of the
ambient air until about sunrise. In fact, it remains within the allowed 1~K
temperature envelope for about 1/2 to 1 hour (summer/winter). This agrees with
experience gained at older (non-evacuated) solar telescopes, where observations
with diffraction-limited quality have been reported for short time periods just
after exposing the primary to sunlight. Of course, once these telescopes and
their optics heated up, the image quality rapidly deteriorated. The temperature
difference between the primary mirror and ambient air $T_{\rm M1} - T_{\rm air}$
exceeds the desired 1~K envelope for most of the observing day. In the morning,
M1 is cooler by up to 6~K in the summer and 4~K in the winter, respectively. In
the late afternoon, the temperature difference reverses sign and reaches values
of 4~K in the summer and 2~K in the winter, respectively. As expected, these
effects are smaller in the winter than in the summer. The lag of the
time-delayed M1 response is about 4 to 6 hours and fine-structures of the air
temperature profile are smoothed out. Importantly, M1 now carries excess heat
into the next observing day, which would make maintaining the temperature
difference in the range of $\pm 1$~K even more difficult. In the following, we
will discuss a more realistic scenario adding solar heating and active cooling
of the primary mirror to our model.

\begin{figure}[t]
\includegraphics[bb=0 0 800 600,width=84mm]{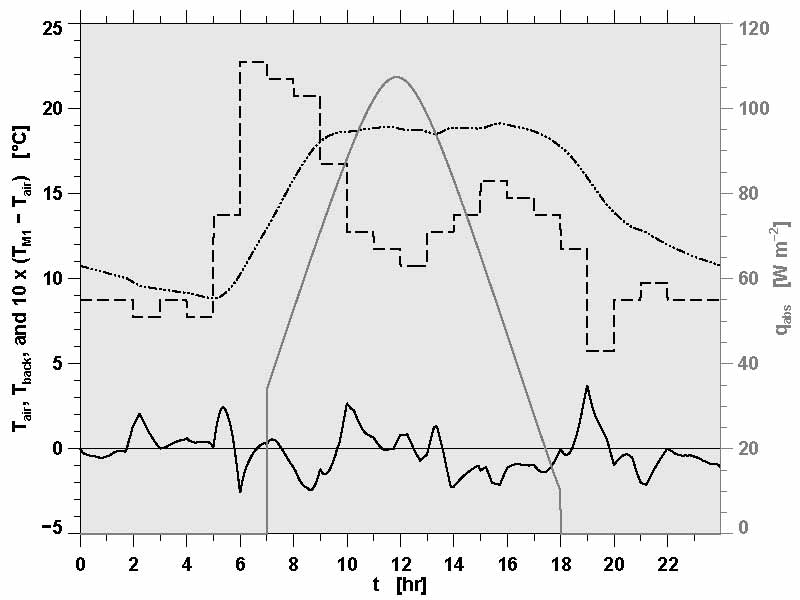}
\hfill
\includegraphics[bb=0 0 800 600,width=84mm]{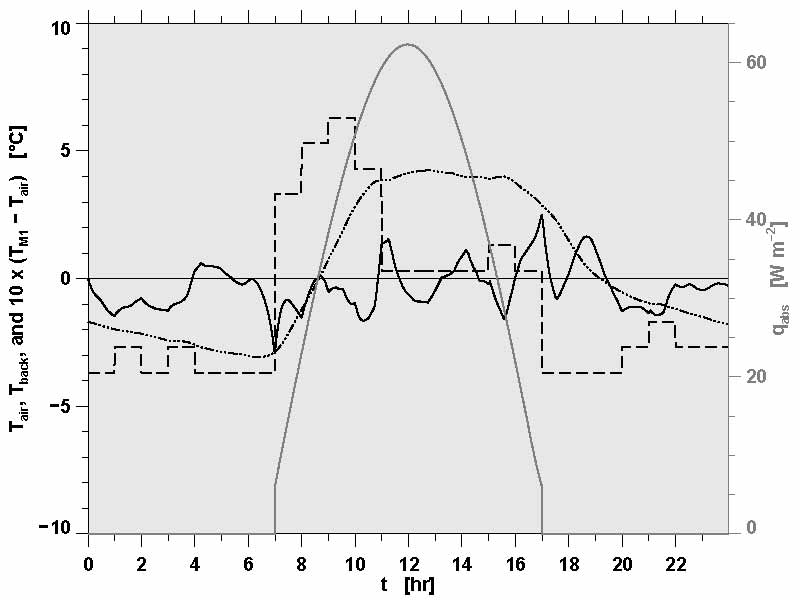}
\caption{{\it Left.} Lump capacity model of the primary mirror M1 with realistic
    temperature profile $T_{\rm air}$ (dashed-triple-dotted) and absorbed solar
    radiation $q_{\rm abs}$ (solid gray) during the summer months. The air
    temperature $T_{\rm back}$ (dashed) to convectively cool the backside of M1
    is changed every hour in multiples of 1~$^{\circ}$C to minimize the
    temperature difference $T_{\rm air} - T_{\rm M1}$ (solid black). The
    temperature difference has been stretched by a factor of 10 to enhance the
    visibility of small-scale temperature variations. {\it Right.} Data for
    winter months (same labeling as in left panel).}
\label{FIG11}
\end{figure}

Starting with the average temperature profiles for summer and winter (see
Figure~\ref{FIG11}), we now add realistic values for the absorbed solar
radiation $q_{\rm abs}$ to the lump capacity model. We assume that the observing
day starts at 7:00~am local time in the morning and ends at 6:00~pm in the
summer and 5:00~pm in the winter (daylight savings time was not taken into
account). Furthermore, the observing days were completely sunny with no cloud
cover. In principle, both radiation and convection could be used as
cooling/heating mechanisms for the primary mirror M1. Since both mechanisms
follow the same formal implementation in the M1 lump capacity model, we only
discuss convective air cooling/heating of M1. In addition, radiative
cooling/heating would require a heat exchanger in close proximity to M1, which
is not possible in the tight confines of the M1 support, which houses the
actuators for the M1 active optics (aO).

Changing the air temperature $T_{\rm back}$ only every hour in multiples of
1~$^{\circ}$C, we are able to keep the primary mirror within a few tenth of
degree Celsius of the ambient air temperature $T_{\rm air}$. The rapid rise of
the ambient air temperature in the morning requires warming M1 by blowing heated
air on the backside. Solar loading by itself is not sufficient to bring M1 fast
enough to the desired temperature. The early morning presents therefore the
greatest challenge for controlling the M1 thermal environment. Since the
backside air temperature $T_{\rm back}$ is almost 15~$^{\circ}$C warmer than the
ambient air temperature $T_{\rm air}$, it is essential the backside of the
primary mirror cell is tightly sealed to avoid an exchange of air. A similar but
inverted control challenge exists in the late afternoon, when M1 has to be
cooled to follow the ambient air temperature. However, the temperature
differentials are not as severe. As expected, the control requirements are much
more relaxed in the winter time, since the day/night temperature difference are
much smaller and so is the solar loading. If a predictive model of the daytime
temperature exists, the thermal control of the primary mirror becomes feasible.
This is certainly the case for sunny days. However, on partially cloudy days
daytime temperature predictions become a challenge and it might not be possible
to keep M1 within the $\pm 1$~$^{\circ}$C operating envelope. On the other hand,
the large clear time fraction at BBSO with more than 300 sunny days, allows to
keep the primary mirror within the temperature margins for most of the time.

One concern of the non-isotropic M1 heating/cooling is that the primary mirror
can become distorted as a result of thermal gradients. This distortion is
proportional to the coefficient of thermal expansion (CTE) of the mirror
material. The CTE of Zerodur is $\alpha = 0.00 \pm 0.10 \times 10^{-6}$~K$^{-1}$
from $0^{\circ} - 50^{\circ}$~C. The major effect is a dominant axial thermal
gradient resulting primarily in a focus error, which can be compensated by
changing the distance between the primary mirror and the secondary mirror
assembly, which is mounted on a hexapod. However, the actual thermal deformation
of the primary mirror will be more complex. This deformation has to be monitored
by a dedicated wavefront sensor, which also measures the  slowly varying
gravitational mirror deformations. These data are fed to a control loop of the
mirror support system.\cite{Yang2006b} The aO support of the primary mirror
consists of 36 actuators in three concentric circles (6, 12, and 18 actuators at
$r_i = 21.8$~cm, 49.1~cm, and 75.1~cm, respectively). Any residual gravitational
or thermal deformations of the primary mirror, which cannot be aO-corrected,
have to compensated by NST's AO system, which is based on the existing BBSO AO
system for the now obsolete 65~cm vacuum reflector.\cite{Denker2007b}

%
%

\section{CONCLUSIONS}

Controlling the thermal environment of the next generation of solar telescopes
will be a major challenge considering the necessarily ``open-design'' of solar
telescopes breaking the 1-meter aperture barrier. We have presented results of
some initial studies to characterize the BBSO site characteristics and the dome
environment in which the future NST will operate. The time-delayed response of
NST's 10-cm thick Zerodur primary mirror to changes of the ambient temperature
requires a detailed understanding of its thermal environment and active measures
to keep the mirror as close to the ambient temperature as possible. For example,
daily temperature predictions become important to determine the optimal
temperature of the primary mirror in the morning. An adaptive scheme to operate
the dome louvers has to be developed, which equalizes the air temperature inside
and outside of the dome, while avoiding wind-shake problems of the optical
support structure, especially for the exposed outrigger that carries the
secondary mirror assembly. Our work has shown that NST can achieve its expected
performance, even in the challenging daytime thermal environment. However, the
results of our studies have to be confirmed and refined in the engineering
first-light phase, which is expected to begin in late 2007. Commissioning of NST
(first-light) is expected about 12 months later.

%
%

\section{ACKNOWLEDGMENTS}

This work was supported by NSF under grants ATM 00-342560, ATM 02-36945, IIS ITR
03-24816 and AST MRI 00-79482 and by NASA under grant NNG0-6GC81G. We would like
to thank Nathan Dalrymple for providing the initial programs to perform the
thermal modeling of the primary mirror and Tom Spirock for providing some of the
BBSO pictures.

%
%


\end{document}